\documentclass[aps,prd,10pt,twocolumn,floatfix,superscriptaddress,preprintnumbers,shownopacs,showkeys,nofootinbib,notoccite,notitlepage]{revtex4-1}

\usepackage{graphicx}
\usepackage{dcolumn}
\usepackage{footmisc}
\usepackage[normalem]{ulem}
\usepackage{xcolor}
\usepackage{cancel}
\usepackage[normalem]{ulem}
\usepackage{bm}
\usepackage{hyperref}
\usepackage{xcolor}
\usepackage{float}
\usepackage{hyperref}
\usepackage{amsmath, amssymb}
\usepackage{tikz-feynman}

\begin{document}
\rightline{FERMILAB-PUB-25-0880-T}
\raggedbottom

\setlength{\abovedisplayskip}{5pt}
\setlength{\belowdisplayskip}{5pt}
\setlength{\abovedisplayskip}{6pt plus 2pt minus 2pt}
\setlength{\belowdisplayskip}{6pt plus 2pt minus 2pt}
\hyphenpenalty=1050

\preprint{}

\title{The Cosmic Neutrino Background is within Reach of Future Neutrino Telescopes}

\author{Gonzalo Herrera}
\email{gonzaloh@mit.edu}
\affiliation{Department of Physics and Kavli Institute for Astrophysics and Space Research, Massachusetts Institute of Technology, Cambridge, MA 02139, USA}
\affiliation{Harvard University, Department of Physics and Laboratory for Particle Physics and Cosmology, Cambridge, MA 02138, USA}
\affiliation{Center for Neutrino Physics, Department of Physics, Virginia Tech, Blacksburg, VA 24061, USA}
\author{Shunsaku Horiuchi}
\email{horiuchi@phys.sci.isct.ac.jp}
\affiliation{Department of Physics, Institute of Science Tokyo, 2-12-1 Ookayama, Meguro-ku, Tokyo 152-8551, Japan}
\affiliation{Center for Neutrino Physics, Department of Physics, Virginia Tech, Blacksburg, VA 24061, USA}
\affiliation{Kavli IPMU (WPI), UTIAS, The University of Tokyo, Kashiwa, Chiba 277-8583, Japan}
\author{Xiaolin Qi}
\email{xiaolinq76@vt.edu}
\author{Ian M. Shoemaker}
\email{shoemaker@vt.edu}
\affiliation{Center for Neutrino Physics, Department of Physics, Virginia Tech, Blacksburg, VA 24061, USA}
\begin{abstract}
The cosmic neutrino background (C$\nu$B) can be boosted to high energies due to scatterings with energetic cosmic rays (CRs) across cosmological scales. Previous calculations focused on neutral current incoherent and coherent elastic scatterings of cosmic-ray protons off relic neutrinos. However, charged current interactions and deep inelastic scatterings are also expected to occur, which enhances the boosted relic neutrino fluxes on Earth. Here, we compute the \textit{total} diffuse boosted cosmic neutrino background (DBC$\nu$B) arising from CRs at all redshifts in the Universe, accounting for neutral current and charged current elastic and deep inelastic scatterings. We find that IceCube already places an upper limit on the cosmic neutrino background overdensity in cosmological scales of ~$\mathcal{O}(100-1000)$ at $E_{\nu}=10^{10}$ GeV, for a lightest neutrino mass of $m_{\nu} \gtrsim 0.1$ eV. We further show that IceCube-Gen2 could test $\mathcal{O}(1-10)$ C$\nu$B overdensities, and the combination of $10$ future neutrino telescopes with similar sensitivity would allow us to test the $\Lambda$CDM expected C$\nu$B density for a lightest neutrino mass compatible with the KATRIN bound.

\begin{figure}[H]
    \centering
\includegraphics[width=0.6\textwidth]{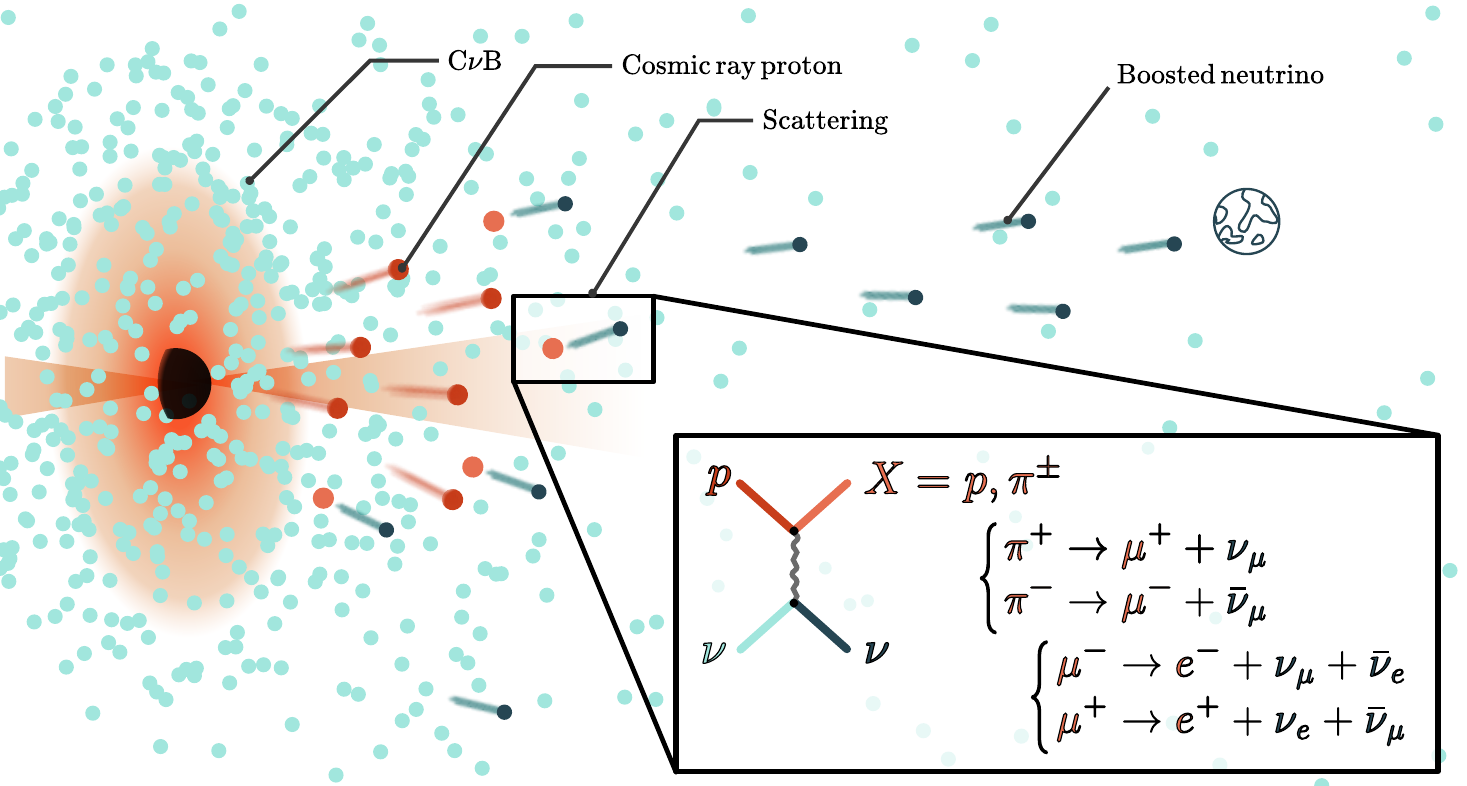}
\caption{Schematic of the cosmic-ray boosted relic neutrino mechanism.}
\label{fig:Diagram_BCnuB}
\end{figure}

\end{abstract}

\maketitle

\textbf{\textit{Introduction.}}
The Cosmic Neutrino Background (C$\nu$B), also known as relic neutrinos, is the last missing piece of the Early Universe thermal history, and a fundamental prediction of our cosmological model, $\Lambda$CDM. It refers to those relic neutrinos that decoupled from the Early Universe plasma about $1s$ after the Big-Bang via weak interaction processes 
\cite{Baumann:2022mni}.

Soon after neutrino decoupling, electron-positron annihilations heated the photon bath via $e^{+}+e^{-} \rightarrow \gamma+\gamma$, which induced a change in the entropy of the system prior- and post-neutrino decoupling of $\left(s_\gamma+s_{\mathrm{e}^{-}}+s_{\mathrm{e}^{+}}\right) / {s_\gamma}=11 / 4$. This change fixed the ratio of temperatures of the Cosmic Microwave Background (CMB) relative to the C$\nu$B. From CMB measurements, it is inferred that the C$\nu$B temperature should be $T_{\nu}=\left(\frac{4}{11}\right)^{1 / 3} T_\gamma \simeq 0.17 \,  \mathrm{meV}$, with total number density today of $n_{\nu}=\frac{3 \zeta(3)}{4 \pi^2} g_{\nu} T_{\nu}^3 \simeq 336 \mathrm{~cm}^{-3}$.

The existence of the C$\nu$B has only been indirectly inferred. $\Lambda$CDM including C$\nu$B predicts the effective number of degrees of freedom $N_{\rm eff}$ at the time of the CMB and the Big Bang Nucleosynthesis (BBN) to be $N_{\mathrm{eff}}^{\mathrm{SM}} \simeq 3.045$ \cite{deSalas:2016ztq, Cielo:2023bqp}, which within 1$\sigma$ deviation is in agreement with CMB and BBN measurements \cite{Planck:2018vyg, Mangano_2011}. These are only sensitive to gravitational effects of the C$\nu$B via its inferred energy density at two specific redshifts.

Despite having indirect evidence for the C$\nu$B, we lack direct evidence. A direct detection is critical, since it could allow us to test neutrino masses and clustering, new physics in the neutrino sector, and most importantly, give us access to the Universe only $1s$ after the Big Bang. 

Yet the task is challenging: the extremely low energy of the C$\nu$B leads to suppressed cross sections in the Standard Model and tiny energy depositions. For instance, consider the coherent scattering of the C$\nu$B off nuclei $N$ at some Earth based detectors. The cross section is roughly $\sigma_{\nu N} \sim E_{\nu}^2 / M_Z^4 \sim 10^{-60} \mathrm{~cm}^2$, with momentum transfers of $q \sim \mathrm{meV}$, which lie far below current detection thresholds and yield event rates $\sim 17$ orders of magnitude smaller than the reach of existing underground detectors, \textit{e.g.}~\cite{Hyper-Kamiokande:2018ofw}.

One way to address the first problem is to consider neutrino capture on a beta-unstable nucleus, which is a reaction with no energy threshold \cite{Weinberg:1962zza}. Currently, KATRIN provides the most stringent sensitivity to this process, placing a limit on the C$\nu$B overdensity of $10^{11}$ on Earth \cite{KATRIN:2022kkv}. Future experiments like PTOLEMY with larger detector masses have been proposed to lower this bound by orders of magnitude \cite{PTOLEMY:2019hkd}. Other proposals have been discussed, with comparable sensitivities to KATRIN \cite{Weiler:1999ny,Domcke:2017aqj,Bauer:2022lri,Brdar:2022kpu,Perez-Gonzalez:2023llw,Bondarenko:2023ukx,Ciscar-Monsalvatje:2024tvm,Chauhan:2024deu,Das:2024thc,Dey:2024agv, delCastillo:2025qnr}, with small enhancements when aided by Beyond the Standard Model interactions, \textit{e.g.}~\cite{Chauhan:2024fas, Dev:2024yrg, Martinez-Mirave:2024dmw, Poddar:2024thb, Franklin:2024amy, Perez-Gonzalez:2023llw, Maitra:2025opp, Lambiase:2025twn, Kaplan:2024ydw, Aliberti:2024udm}.

Furthermore, it has been discussed that the detection of the C$\nu$B with beta decay experiments suffers from the Heisenberg uncertainty principle, which may forbid distinguishing the endpoint of the beta decay electron from the neutrino-induced peak \cite{PTOLEMY:2022ldz, Cheipesh:2021fmg, Nussinov:2021zrj}. In particular, Ref.~\cite{Cheipesh:2021fmg} demonstrated that the energy uncertainty on the beta decay electrons may be larger than the neutrino mass, impeding the resolution of the C$\nu$B peak.

A perhaps more promising idea to detect the C$\nu$B has recently been discussed in a series of works \cite{Ciscar-Monsalvatje:2024tvm, DeMarchi:2024zer, Herrera:2024upj, Zhang:2025rqh}. Energetic cosmic rays (CR) can scatter off the C$\nu$B via Standard Model processes, boosting it to high-energies, see Fig.\ref{fig:Diagram_BCnuB} for a schematic of this mechanism. The consideration of this process within the Milky Way and extragalactic CR accelerators allowed to place overdensity limits comparable to that from KATRIN, but on larger physical scales \cite{Ciscar-Monsalvatje:2024tvm}. Ref.~\cite{Herrera:2024upj} extended the idea to neutral current elastic scatterings occurring across different redshifts, finding further enhancements in the boosted neutrino fluxes on Earth by orders of magnitude compared to the local galactic contribution. 

In this paper, we include contributions from charged current interactions and deep inelastic scattering processes which were neglected in previous studies. We then confront our predictions with the sensitivity of current and future (ultra)high-energy neutrino experiments, and with expectations on the C$\nu$B overdensity from Pauli exclusion principle, gravitational neutrino clustering, and clustering in Beyond the Standard Model scenarios. 


\vspace{6pt}
\textbf{\textit{Diffuse Boosted Cosmic Neutrino Background.}} The diffuse flux of C$\nu$B boosted by CR protons across all redshifts can be expressed in full generality as \cite{Herrera:2024upj}
\begin{equation}
\begin{aligned}
\frac{d \phi_{\nu}}{d T_{\nu}} &= \int_{z_{\text{min}}}^{z_{\text{max}}} dz \, \frac{c}{H_0} \frac{1}{\sqrt{(1+z)^3 \Omega_m + \Omega_{\Lambda}}} \, f_i(z) \, \eta \, n_\nu (1+z)^3 \\
&\times \int_{T_{p}^{\rm min}}^{\infty} d T_p \, \frac{d\sigma_{p\nu}}{dT_{\nu}} \, \frac{d \phi_p}{d T_p} \, \Theta\left[T_\nu^{\max}(T_p) - T_\nu(1+z)\right],
\label{eq:flux}
\end{aligned}
\end{equation}
where $T_p$ is the proton kinetic energy, $f_i(z)$ is the redshift evolution of the CR flux, $n_{\nu}=336$ cm$^{-3}$ is the average density of the C$\nu$B in the Universe today, which we evolve in redshift with the standard $(1+z)^3$, ${d\sigma_{p\nu}}/{dT_{\nu}}$ is the differential proton-neutrino scattering cross section, $d\phi_p/dT_p$ is the CR spectrum, and $T_\nu^{\max }$ is the maximum kinetic energy transferred to a neutrino in a single collision. We adopt $H_0=67.4 \mathrm{~km} / \mathrm{sec} / \mathrm{Mpc}, \Omega_{\Lambda}=0.685$, and $\Omega_m=0.315$ \cite{Planck:2018vyg}. The CR energy threshold $T_{p}^{\rm min}$ varies for different interaction channels.
\begin{figure*}[t!]
    \centering
    \includegraphics[width=0.48\textwidth]{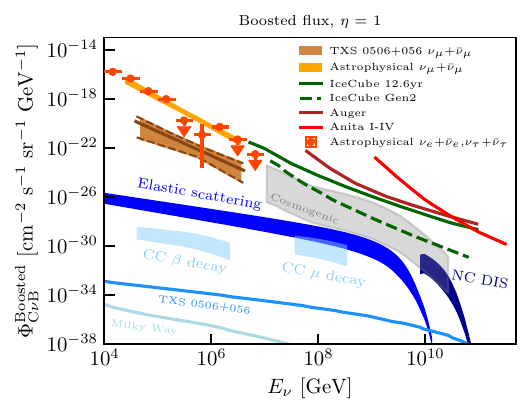}
    \includegraphics[width=0.494\textwidth]{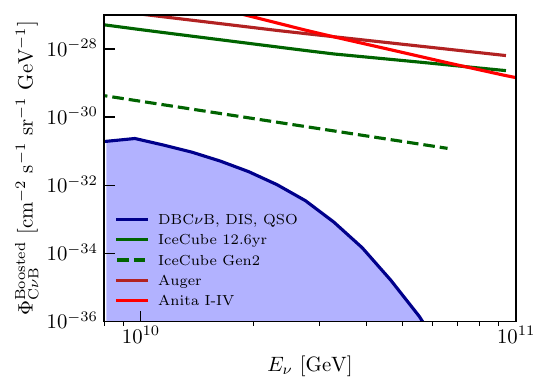}
    
    \caption{\textit{Left panel:} The C$\nu$B flux boosted by CR protons via different scattering channels, summed over all three flavors. The light blue bands indicate the contributions from CC quasi-elastic interactions, via either $\beta$ decay or $\mu^{\pm}$ decay. The blue colored band corresponds to the NC elastic scattering channel, already computed in Ref. \cite{Herrera:2024upj}. The dark blue band corresponds to our NC DIS contribution. The thickness of the bands reflects the uncertainty in the CR source evolution. In solid blue lines, we further show the boosted C$\nu$B fluxes from the Milky Way and TXS 0506+056 \cite{Ciscar-Monsalvatje:2024tvm}. All predictions shown assume a neutrino mass of $m_\nu=0.1$ eV. For comparison, we present the total flux of atmospheric neutrinos and diffuse extragalactic neutrinos measured by terrestrial experiments \cite{Super-Kamiokande:2015qek, IceCube:2016umi, IceCube:2020acn}, along with the current upper limits on ultra-high-energy neutrinos from IceCube (green, \cite{IceCubeCollaborationSS:2025jbi}), Auger (dark red, \cite{PierreAuger:2015ihf}), and ANITA I-IV (red, \cite{ANITA:2018, ANITA:2019wyx}). Sensitivity projections for IceCube-Gen2 radio (10 yr) (dotted green, \cite{IceCube-Gen2:2021rkf}) and the expected range of cosmogenic-neutrino fluxes (grey band, \cite{Fang:2017zjf,GRAND:2018iaj}) are also displayed. \textit{Right panel}: Same as left panel, but zoomed in the NC DIS part. In blue shading is the ultra-high-energy ``neutrino fog" induced by the deep-inelastically boosted C$\nu$B (labeled DBC$\nu$B, DIS, QSO). The upper end of the blue colored region corresponds to CRs following the QSO evolution function.}
    \label{fig:flux_everything}
\end{figure*}

Since the CR flux ${d \phi_{p}}/{d T_p}$ beyond the Earth's vicinity cannot be directly measured, we adopt a broken power-law spectrum for simplicity. For energies below $E_{p}=10^{7}$ GeV we adopt a slope of $-2.7$ which is consistent with CRs measured on Earth. Above $E_{p}=10^{7}$ GeV, we adopt power-law spectra $\propto E^{-\alpha}$ where following Ref.~\cite{Kotera_2010}, the slope $\alpha$ is source dependent and is determined by requiring the CR spectrum after propagation through the CMB to match the observed spectrum on Earth. Since the sources of CRs at the highest energies are still uncertain (see, \textit{e.g.}, \cite{AlvesBatista:2019tlv}), we follow previous studies and consider several source classes. Accordingly, we model the CR source redshift evolution as, 
\begin{equation}
f_i(z) = \frac{N_i \left(z\right)}{N_i\left(z_{\rm min}\right)},
\end{equation}
where $N_i(z)$ is the CR source distribution function at different redshifts, normalized such that $f_{i}(z_{min}) = 1$. We consider three possible CR sources with well-studied distributions: the cosmic star-formation rate (SFR), the Fanaroff–Riley II or quasar (QSO) distribution, and the gamma-ray burst (GRB) rate. These have been obtained from a combination of observations and theoretical inputs, and can be expressed analytically (e.g., \cite{Kotera_2010}). Details can be found in the Appendix \ref{sec:CR_evolution}. Due to the effects of energy redshifting, the choice results in determining $\alpha$: for the SFR, $\alpha=2.5$; for the QSO, $\alpha=2.3$; and for the GRB rate, $\alpha=2.4$ (see Ref.~\cite{Kotera_2010} for details). For normalization, we adopt $z_{\rm min}=2.37 \times 10^{-6}$ which is approximately equal to the size of the Milky Way galaxy, $ \simeq 10$ kpc. We nominally take $z_{\rm max}=6$ for all sources, and assume CRs to consist purely of protons.

\vspace{6pt}
\textbf{\textit{Neutral current elastic scatterings.}} For center of mass energies below $\sqrt{s} \simeq 1$ GeV, CRs scatter off the C$\nu$B via neutral current (NC) elastic scatterings (ES), i.e. $\nu+p \rightarrow \nu+p$, yielding a boosted neutrino and a degraded CR proton in the final state. The cosmological boosted C$\nu$B flux arising from this channel has been previously discussed \cite{Herrera:2024upj,Zhang:2025rqh}. A description of the ES cross section is provided in the Appendix \ref{sec:cross_sections}.
In Fig.~\ref{fig:flux_everything}, the ES boosted C$\nu $B is shown as the blue band. The band width reflects the uncertainties from different CR source distributions: the upper bound, corresponding to QSO evolution, represents the most optimistic prediction, whereas the lower bound, given by SFR evolution, provides the most conservative estimate. The C$\nu$B boosted by single galaxies, \textit{e.g.}, the Milky Way or TXS 0506+056\cite{Ciscar-Monsalvatje:2024tvm} are also shown. Compared to these single-galaxy cases, including the contribution from all galaxies leads to an enhancement of several orders of magnitude, underscoring the importance of accounting for the full population.  

\vspace{6pt}
\textbf{\textit{Charged current interactions.}} In the quasi-elastic scattering regime, the following charged current (CC) processes can take place for each antineutrino flavor $\bar{\nu}_{\ell}$,
\begin{align}
p+\bar{\nu}_{\ell} \rightarrow n+\ell^{+},
\end{align}
where $\ell^{+}$ is the outgoing positron $\left(e^{+}\right)$, antimuon $\left(\mu^{+}\right)$, or antitau $\left(\tau^{+}\right)$. In the kinematically allowed regime, the cross section for this process is approximately 2.394 times that of the NC elastic interactions \cite{formaggio2012ev}. CC processes can yield neutrinos in the final-state from both neutron decay and charged lepton decay. A description of the kinematics and corresponding final neutrino energies can be found in the Appendix \ref{sec:CC_contribution}.

The boosted flux arising from CC processes is displayed in Fig.~\ref{fig:flux_everything} as light blue bands. Compared to the NC boosted flux, the CC boosted fluxes contribute only marginally. This is expected: in CC interactions, the boosted neutrinos originate from the decay of secondary leptons, whereas in NC interactions the target neutrinos are directly boosted, carrying a larger fraction of the incoming CR energy. Thus, for boosted neutrinos of the same energy, CC processes require higher-energy CR protons to scatter off the $C\nu$B. Since the CR spectrum follows a broken power law that falls steeply, the flux of boosted neutrinos from CC processes is suppressed compared to NC processes. 

\begin{figure*}[t!]
    \centering
    \includegraphics[width=0.48\textwidth]{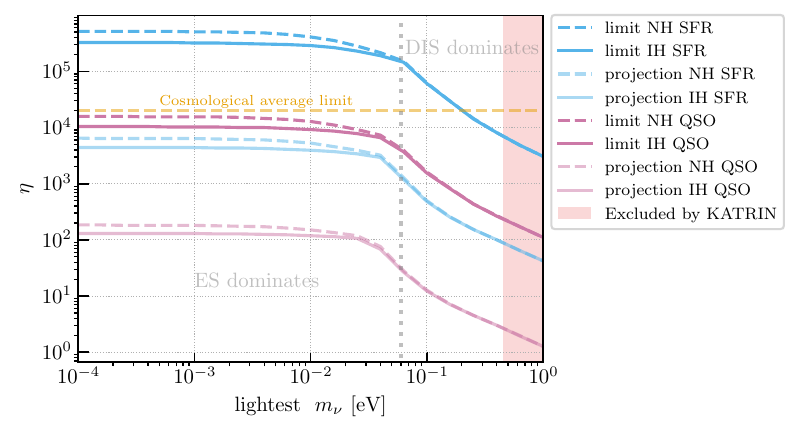}
    \includegraphics[width=0.50\textwidth]{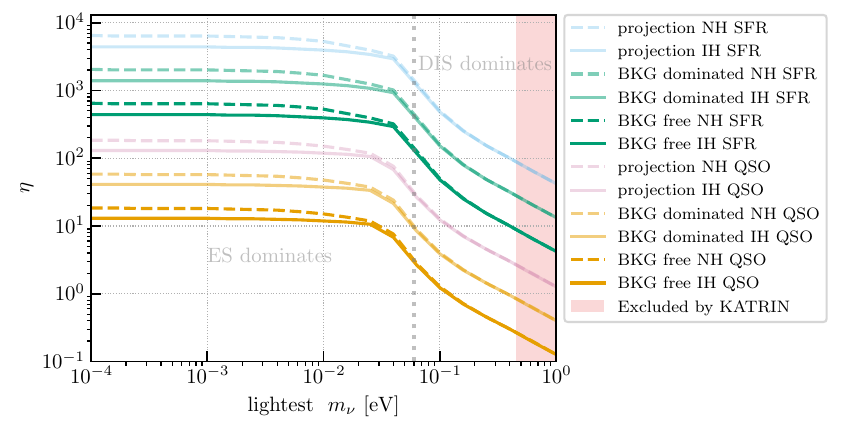}
    \caption{\textit{Left panel}: Current and projected upper limits on the C$\nu$B overdensity ($\eta$) versus lightest neutrino mass, for different CR models and normal (inverted) neutrino mass hierarchies in solid (dashed). The best limit is obtained from a combination of the elastic and deep-inelastic scatterings. For neutrino masses to the left (right) of the dotted vertical gray line, elastic (deep-inelastic) scattering drives the most stringent limit. Limits set by NC elastic and NC DIS processes respectively can be found in the Appendix \ref{sec:ES_and_DIS_limits}. The dashed orange line denotes the maximum $\eta$ allowed by Pauli principle on cosmological scales \cite{Bondarenko:2023ukx} (see Appendix \ref{sec:neutrino_overdensity}). The upper bound on the lightest neutrino mass from KATRIN is shown as a shaded red region \cite{KATRIN:2024cdt}. \textit{Right panel}: Projected upper limits on the C$\nu$B from a combination of 10 future experiments with comparable sensitivity at ultra-high energies to IceCube-Gen2, both for a background-free scenario (dubbed ``BKG free"), and for a background-dominated scenario (dubbed ``BKG dominated). For comparison, we also show the IceCube-Gen2 only projected sensitivity.}
    \label{fig:eta_single}
\end{figure*}

\vspace{6pt}
\textbf{\textit{Deep inelastic scattering.}}
Ultra-high-energy cosmic rays (UHECRs) scattering off relic neutrinos can reach center-of-mass energies in the deep inelastic scattering (DIS) regime. For instance, a GZK proton of energy $E_{\rm GZK} \simeq 5 \times 10^{19}$ eV scattering off a neutrino with mass $m_{\nu}=0.1$ eV yields a center of mass energy of $\sqrt{s} \simeq \sqrt{2m_{\nu}E_{\rm GZK}} \simeq 4.6$ GeV, \textit{i.e.,} in the DIS regime.

A description of the DIS cross section is given in the Appendix \ref{sec:cross_sections}. In our calculation, DIS refers to the directly boosted neutrinos from neutral current DIS (NC DIS). Specifically, charged current DIS and neutrinos from the further decay of secondary hadrons are not considered. This is reasonable given these neutrinos will not dominate for the same reasons as discussed for the CC interactions.

DIS occurs only for proton energy above a certain threshold. We estimate the threshold by requiring the corresponding neutrino energy in the proton rest frame to be $\gtrsim 2 \, \mathrm{GeV}$ \cite{formaggio2012ev},
\begin{align}
E_p^{\mathrm{th, DIS}} \simeq \min\{2\mathrm{GeV} \times \frac{m_p}{m_{\nu}},\; 10^{11}\mathrm{GeV}\}, \label{E_p_thred_DIS}
\end{align}
where $10^{11}\mathrm{GeV}$ is maximal proton energy considered in this work. For an effective neutrino mass $m_{\nu} \simeq 0.1$ eV, we get $E_p^{\mathrm{th, DIS}} \simeq 1.876 \times 10^{10} \mathrm{GeV}$. 

Above the threshold, the elastic contributions from NC and CC channels are subdominant compared to DIS. The DIS boosted neutrino flux is shown in the left panel of Fig.~\ref{fig:flux_everything} as the dark blue band (zoomed in the right panel). The neutrino mass is fixed to $m_{\nu} = 0.1$ eV. For comparison, several upper limits on cosmogenic neutrinos and projected limits by IceCube Gen2 are shown. If the CR evolution tracks the QSO evolution, the DIS boosted neutrino flux is only $\sim$ 1 order of magnitude away from detectability at IceCube-Gen2.


\vspace{6pt}
\textbf{\textit{Limits from current neutrino telescopes.}}
Neutrino telescopes have placed limits on the ultra-high energy neutrino flux. If the neutrino overdensity $\eta$ is large enough, the resulting boosted $C\nu$B will have a flux that may exceed these limits, usually reported at $90\%$C.L. Thus, neutrino telescopes place an upper bound on the C$\nu$B overdensity by requiring that the boosted flux does not exceed any of the existing limits. In this work, we consider current upper limits from IceCube 12.6yr \cite{IceCubeCollaborationSS:2025jbi}, Auger \cite{PierreAuger:2015ihf}, and Anita I-IV \cite{ANITA:2019wyx}.  

In the left panel of Fig.~\ref{fig:eta_single}, current limits on $\eta$ are shown in solid (dashed) lines for inverted (normal) neutrino mass hierarchy as a function of the lightest neutrino mass. The dependence on the CR evolution is manifested by different colors. The limits fall somewhat sharper with neutrino mass than the NC ES case, since for larger neutrino masses, the threshold energy $E_p^{\mathrm{th, DIS}}$ in Eq.~\eqref{E_p_thred_DIS} decreases, thereby extending the proton energy range that contributes to the DIS boosted flux, leading to a significant enhancement.

\vspace{6pt} 
\textbf{\textit{Projected limits from future neutrino telescopes.}}
We further derive future detection prospects with IceCube-Gen2 \cite{IceCube-Gen2:2020qha, IceCube-Gen2:2021rkf}, and from a combination of 10 future neutrino experiments with comparable sensitivities. Several experiments are expected to operate at ultra-high energies, and we refer the reader to the Appendix ~\ref{sec:future_telescopes} for a brief description.

In the right panel of Fig.~\ref{fig:eta_single}, we show future sensitivity projections to the C$\nu$B. The limits dubbed ``projection" correspond to IceCube-Gen2-only. 

We further combine independent, background-limited limits on a common flux normalization via inverse-variance weighting
\(\sigma_{\rm comb}^{-2}=\sum_i \sigma_i^{-2}\). For comparable experiments (\(\sigma_i\!\approx\!\sigma\)), the 90\%\,C.L.~upper limit improves as \(\mathrm{UL}_{\rm comb}\!\simeq\!\mathrm{UL}/\sqrt{N}\). In the background-free Poisson regime, limits scale with total exposure instead. In the absence of a cosmogenic neutrino signal, these experiments are expected to remain background-free, while, if the cosmogenic neutrino flux is detected, these experiments would be background-dominated when searching for the C$\nu$B. Therefore, we show limits for both scenarios. For a combination of 10 background free experiments with sensitivities comparable to that expected from IceCube-Gen2, the C$\nu$B density expected in $\Lambda$CDM could be tested at $90\%$ C.L., for a lightest neutrino mass of $m_{\nu} \gtrsim 0.1$ eV.

\vspace{6pt}
\textbf{\textit{Conclusions.}} 
Detecting the C$\nu$B is a crucial goal for astroparticle physics and cosmology. It has recently been discussed that the scatterings of CRs with C$\nu$B over cosmological scales induce a component of (ultra)high-energy boosted C$\nu$B that may be detected with neutrino telescopes. Here, we included the contributions from NC DIS processes and CC quasi-elastic scatterings on the ensuing neutrino fluxes, finding that the former significantly enhances the sensitivity of experiments at energies $E_{\nu} \gtrsim 7 \times 10^{9}$GeV, while the latter is subdominant.

When including the NC DIS scattering channel, we find that IceCube already places an upper limit on the C$\nu$B overdensity in cosmological scales of ~$\mathcal{O}(100-1000)$ at $E_{\nu}\simeq10^{10}$ GeV, for a lightest neutrino mass of $m_{\nu} \gtrsim 0.1$ eV. We further derived sensitivity projections with IceCube-Gen2, showing that $\mathcal{O}(1-10)$ C$\nu$B overdensities for a lightest neutrino mass of $m_{\nu} \gtrsim 0.1$ eV could be tested, and the combination of 10 future neutrino telescopes with similar sensitivity would allow a test of the $\Lambda$CDM expected C$\nu$B density for a lightest neutrino mass compatible with laboratory bounds from KATRIN.

Importantly, we show that IceCube is able to place constraints on the C$\nu$B that exceed the theoretical limit obtained from Pauli exclusion principle on cosmological scales. To the best of our knowledge, this is the only probe of the C$\nu$B that is able to surpass this fundamental limit. The CR boosted C$\nu$B also constitutes an unavoidable ``fog" for other contributions of Beyond the Standard Model physics to the (ultra)high-energy neutrino flux, in a similar fashion to the solar neutrino ``fog" at dark matter direct detection searches.

Our work could be improved in different directions. For instance, we did not include contributions from $\tau$ production in CC scatterings, which would further enhance the neutrino fluxes on Earth. Along these lines, it should be pointed out that the neutrino flavor composition from CR boosted relic neutrinos differs from that expected from cosmogenic neutrinos, which may offer an additional discrimination channel: cosmogenic neutrinos from pion decay are produced with an initial ratio $\left(\nu_e: \nu_\mu: \nu_\tau\right)=1:2:0$, with oscillations averaging to a detectable democratic ratio of $1:1:1$. In contrast, the boosted C$\nu$B neutrinos originate from mass eigenstates, so their detected flavor ratio composition is determined by the relative upscattered $\nu_i$ fluxes and their PMNS projections, which for normal and inverted ordering may not be democratic.
Furthermore, a proper comparison of the spectral dependence of the boosted C$\nu$B fluxes with the cosmogenic neutrino flux and other potential neutrino backgrounds is needed for future detection. Finally, calculations incorporating a CR composition beyond purely protons will allow a cleaner comparison, as heavy nuclei may suppress the cosmogenic neutrino flux more strongly than it suppresses the boosted C$\nu$B.
\medskip

\textbf{\textit{Acknowledgments.}}
We thank Mar Císcar Monsalvatje for discussions and help with Fig.~\ref{fig:Diagram_BCnuB}. We are grateful to Carlos Argüelles and Nick Kamp for useful feedback. The work of GH is supported by the Neutrino Theory Network Fellowship with contract number 726844. The work of SH is supported by NSF Grant No.~PHY-2209420 and JSPS KAKENHI Grant Number JP22K03630 and JP23H04899. IMS and XQ are supported by the U.S. Department of Energy under the award number DE-SC0020262. This manuscript has been authored by FermiForward Discovery Group, LLC under Contract No. 89243024CSC000002 with the U.S. Department of Energy, Office of Science, Office of High Energy Physics. This work was supported by World Premier International Research Center Initiative (WPI Initiative), MEXT, Japan.

\bibliography{References}

\clearpage
\appendix
\onecolumngrid

\section{Cosmic ray evolution}\label{sec:CR_evolution}

The flux of CR boosted relic neutrinos depends on the CR flux evolution over redshift \footnote{We note that the CR evolution functions presented in Ref.~\cite{Herrera:2024upj} require corrections. The corrected forms are given in this appendix. These changes do not affect the numerical results or conclusions of Ref.~\cite{Herrera:2024upj}.}. For the SFR evolution, we take \cite{Hopkins:2006bw},
\begin{equation}
\log_{10}N_{\mathrm{SFR}}(z) = \begin{cases}
    a \log_{10} (1+z) + b & z \leq 1.04 \\
    c \log_{10} (1+z) + d & 1.04 < z \leq 4.48  \\
    e \log_{10} (1+z) + f & 4.48 < z \leq 6
    \label{eq: SFR_CR}
\end{cases}
\end{equation}
with $a=3.28, b=-1.82, c=-0.26, d=-0.724, e=-8.0$, and $f=4.99$. For the QSO evolution, we consider the source density from Ref.~\cite{Wall:2004tg}
\begin{equation}
\log_{10}N_{\rm QSO}(z)=-a_0+a_1 z-a_2 z^2+a_3 z^3-a_4 z^4,
\label{eq: QSO_CR}
\end{equation}
with $a_0$=12.49, $a_1$=2.704, $a_2$=1.145, $a_3$=0.1796, and $a_4$=0.01019. For the GRB distribution function, we use the parametrization from \cite{Le:2006pt},
\begin{equation}
N_{\rm GRB}(z)=\frac{1+a_2 z/a_1}{1+(z/a_3)^{a_4}},
\label{eq: GRB_CR}
\end{equation}
with the lower values corresponding to $a_1=0.015$, $a_2=0.12$, $a_3=3.0$, and $a_4 = 1.3$, and the upper values corresponding to $a_1=0.011$, $a_2=0.12$, $a_3 = 3.0$, and $a_4=0.5$.

\begin{figure}[H]
    \centering
\includegraphics[width=0.55\textwidth]{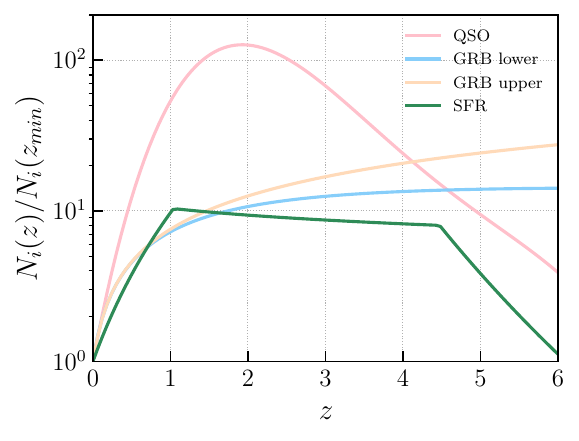}
\caption{Normalised Cosmic ray evolutions considered in this paper: SFR, GRB lower, GRB upper, and QSO. $N_{i}(z)$ are given by Eq.~(\ref{eq: SFR_CR})-(\ref{eq: GRB_CR}).}
\label{fig:CR_evol}
\end{figure}

\section{Proton--neutrino cross sections across energy}\label{sec:cross_sections}
For neutral current (NC) proton-neutrino scatterings in the elastic regime ($\nu+p \rightarrow \nu+p$), the differential cross section summed over $\nu$ and $\bar{\nu}$ reads \cite{Schmitz1997Neutrinophysik,Giunti:2007ry,DeMarchi:2024zer}
\begin{equation} \label{eq:cross_section_el}
\frac{d \sigma_{\nu p}^{\mathrm{el}}}{d E_\nu}=\frac{2 G_F^2 m_\nu m_p^4}{\pi\left(s-m_p^2\right)^2}\left[A\left(Q^2\right) \pm B\left(Q^2\right) \frac{s-u}{m_p^2}+C\left(Q^2\right) \frac{(s-u)^2}{m_p^4}\right],
\end{equation}
where $E_{\nu}$ is the outgoing neutrino energy, $G_F$ is the Fermi constant, $s$ (also known as the invariant mass) and $u$ are the two Mandelstam variables of the system, and $Q^2 = 2m_{\nu}(E_{\nu} - m_{\nu})$ is the squared momentum transfer between incoming and outgoing neutrinos. The three coefficients A, B, C are given by
\allowdisplaybreaks
\begin{gather}
A_N(q^2)=\frac{q^2}{m_N^2}\Bigg[\left(1+\frac{q^2}{4 m_N^2}\right)\!\big(G_A^{ZN}\big)^2
-\left(1-\frac{q^2}{4 m_N^2}\right)\!\left(\big(F_1^{ZN}\big)^2-\frac{q^2}{4 m_N^2}\big(F_2^{ZN}\big)^2\right)
+\frac{q^2}{m_N^2}\,F_1^{ZN} F_2^{ZN}\Bigg], \\
B_N(q^2)=\frac{q^2}{m_N^2}\, G_A^{ZN}\left(F_1^{ZN}+F_2^{ZN}\right), \\
C_N(q^2)=\frac{1}{4}\left[\big(G_A^{ZN}\big)^2+\big(F_1^{ZN}\big)^2+\frac{q^2}{4 m_N^2}\big(F_2^{ZN}\big)^2\right],
\end{gather}
where positive/negative sign applies to neutrinos/antineutrinos. For charged current, we multiply Eq.~(\ref{eq:cross_section_el}) by a factor of 2.39 \cite{formaggio2012ev}. Finally, for neutral current deep inelastic scatterings (NC DIS), the proton-neutrino scattering differential cross section reads \cite{Giunti:2007ry,DeMarchi:2024zer}

\begin{align}
\frac{d \sigma_{\nu N}^{\mathrm{DIS}}}{d E_\nu} \simeq & \sum_{a=q, \bar{q}} \frac{G_F^2\left[\left(g_V^a\right)^2+\left(g_A^a\right)^2\right]}{2 \pi E_N} \nonumber\int_{y_{\min }}^1 \frac{d y}{y^2} \frac{Q^2 f_a^N\left(x, Q^2\right)}{\left[1+Q^2 / M_Z^2\right]^2} g\left(y, Q^2, m_N\right),
\end{align}
where $E_{\nu}$ is the outgoing neutrino energy, $E_{N}$ is the nucleon energy, $M_Z(N)$ is the $Z$ boson (nucleon) mass; $g\left(y, Q^2, m_N\right) \equiv\left(y^2-2 y+2-2 m_N^2 x^2 y^2 / Q^2\right)$, where $y$ is the inelasticity parameter satisfying $y_{\min }=\left(E_\nu-m_\nu\right) / E_N \lesssim y \leq 1$, and $x=\left(E_\nu-m_\nu\right) /\left(E_N y\right)$ is the Bjorken scaling variable;  $g_{V(A)}^a$ is the NC vector (axial) coupling for the quark $a = u, d, s, c, b$ (top quark contribution is neglected); $f_a^N\left(x, Q^2\right)$ is the parton distribution function (PDF) for $a$. In our calculation, PDFs are taken from $\texttt{ManeParse}$, where including the PDF uncertainties only has a negligible impact of $\sim 2\%$ on the boosted flux.

\section{Charged current interactions}\label{sec:CC_contribution}

In this appendix, we explain how the boosted C$\nu$B is obtained via charged current (CC) processes,
\begin{align}
p+\bar{\nu}_{\ell} \rightarrow n+\ell^{+}.
\end{align}
These processes only occur for CR proton energies above a given threshold. In the frame of reference where the non-boosted C$\nu$B is at rest, the calculation is straightforward,
\begin{equation}
    E_{p}^{\mathrm{th}, \nu} = m_{n} + m_{l^{+}},
\end{equation}
where $m_X$ denotes the rest mass of $X$. Then, from the Lorentz invariance of Mandelstam variable $s$, the energy threshold in the lab frame $E_p^{th}$ is obtained,
\begin{equation}
E_p^{\mathrm{th}}=\frac{\left(m_n+m_{\ell}\right)^2-m_p^2}{2 m_{\bar{\nu}}}.
\end{equation}

For simplification, let us consider an effective neutrino mass of $m_{\nu} \simeq 0.1$ eV for all flavors. This gives us an estimate of the necessary incoming proton energy for such CC interactions to occur,
\begin{align}
E_p^{\mathrm{th}, e^{+}} \simeq 2.2 \times 10^{7} \mathrm{GeV} \\
E_p^{\mathrm{th }, \mu^{+}} \simeq 1.1 \times 10^{9} \mathrm{GeV}\\
E_p^{\mathrm{th }, \tau^{+}} \simeq 3.2 \times 10^{10} \mathrm{GeV}.
\end{align}
We can see that the production of positrons and antimuons occurs in a wide range of CR proton energies considered in this work, while the antitau production only occurs at the largest CR proton energies, close to the GZK limit. By comparing $E_p^{\mathrm{th }, \tau^{+}}$ to $E_p^{\mathrm{th }, \rm DIS}$ in Eq.~\eqref{E_p_thred_DIS}, we notice that antitau production occurs when the incoming protons are already energetic enough to trigger deep inelastic scatterings. Therefore, this channel can be neglected to a good approximation.

These processes as stated simply ``erase" the cosmic neutrino background. However, the resulting neutron can undergo beta decay, yielding additional antineutrinos,
\begin{equation}
n \rightarrow p+e^{-}+\bar{\nu}_e.
\end{equation}
The resulting average energy of antineutrinos from a given initial proton energy $E_p$ is \cite{Moharana:2011hh}
 \begin{equation}
E_{\bar{\nu_{e}}}=\frac{(m_n-m_p-m_e)}{m_n} \times \, E_{n},
 \end{equation}
where $E_{n}$ is the neutron energy. To estimate $E_{n}$, we assume that $\nu_{e}$ and $\nu_{\mu}$ scattering off protons produce the same number of neutrons. At the lower energy threshold where the CC interaction processes become relevant, $E_{n}$ could be fully determined to be approximately $m_n$. When the C$\nu$B has an energy of $T_{\nu} = 10 \, \mathrm{GeV}$ in the proton rest frame, the resulting neutron will have an energy of $E_n \simeq 0.45 \times T_{\nu}$ \cite{Gandhi:1995tf}. Within the energy range relevant for CC interactions, we interpolate linearly among these two kinematical regimes.

It should be noted that the additional proton from the beta decay can be reprocessed, further scattering off the cosmic neutrino background. For simplicity, we restrict ourselves to compute the boosted neutrino flux arising from the first scattering process, which is conservative.

In addition to neutron decay, for the charged current interaction proton scattering off $\bar{\nu}_{\mu}$, further production of neutrinos can arise from antimuon decay,
\begin{equation}
\mu^{+} \rightarrow  e^{+} \nu_e \bar{\nu_{\mu}}.
\end{equation}
For our purposes, the antimuon is typically highly boosted compared to the positron mass; therefore, we assume that the decay is roughly isotropic in energy,
\begin{equation}
E_{\nu} \simeq \frac{E_{\bar{\mu}}}{3},
\end{equation}
where $E_{\bar{\mu}}$ could be approximated similarly as $E_n$.

\section{Overdensity limits from ES and DIS processes}\label{sec:ES_and_DIS_limits}

In this appendix, we provide the current and projected sensitivity limits set by only considering elastic scattering (ES) and neutral current deep inelastic scattering (NC DIS), respectively, for a better comparison.  

\begin{figure}[H]
		\centering
        \includegraphics[width=0.49\textwidth]{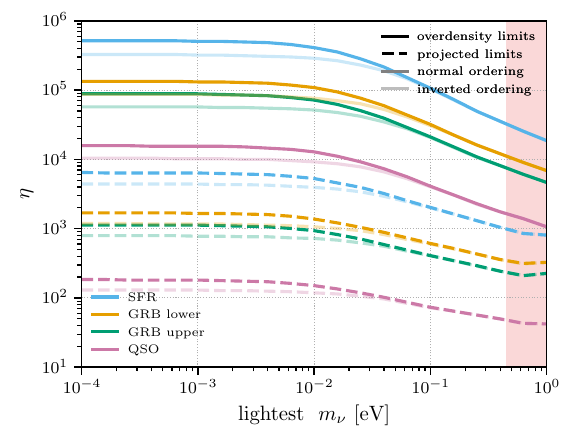}
        \includegraphics[width=0.49\textwidth]{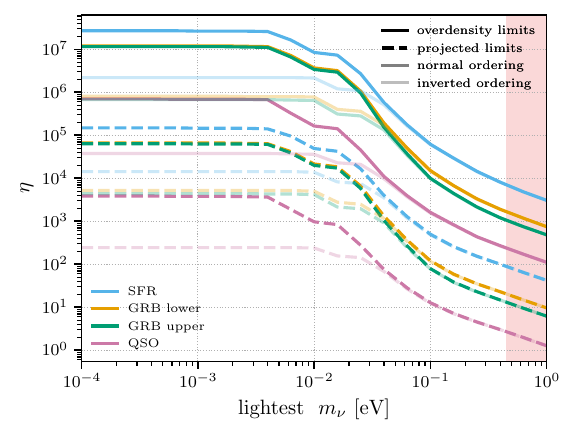}
		\caption{\textit{Left panel}: Current and projected limits set by ES processes. Different colors represent different CR source evolutions. Solid lines are current limits, dashed lines are projected limits. Opaque lines indicate normal neutrino mass ordering, and translucent ones indicate inverted ordering; \textit{Right panel}: Same as left panel, but with DIS processes considered.}
        \label{fig:ES_DIS_etas}
\end{figure}

Left panel in Fig.~\ref{fig:ES_DIS_etas} shows the ES limits, there is a decrease of $O(1)$ in $\eta$ when neutrinos become heavier, consistent with the results obtained in \cite{Ciscar-Monsalvatje:2024tvm}. For smaller neutrino masses, ES provides a more stringent limit than DIS. The right panel corresponds to NC DIS limits, which change more quickly when neutrinos become heavier, due to the reason presented in the main text. For a neutrino mass of $m_{\nu}\gtrsim 6\times10^{-2}\mathrm{eV}$, DIS processes result in stronger limits than those obtained with ES.

\section{Future neutrino telescopes}\label{sec:future_telescopes}

The advent of several future neutrino telescopes will allow for the possibility to combine their data to enhance the sensitivity to the cosmic neutrino background. Some of the benefits arising from a global network of neutrino telescopes were discussed in \cite{Schumacher:2025qca}. Here we provide a brief summary of the landscape of future neutrino telescopes, justifying our choice of $N=10$ when deriving projections for the C$\nu$B in the main text. 

\begin{itemize}

\item[$\bullet$] \textbf{IceCube-Gen2:} The planned extension of IceCube will expand the optical array to enhance the \(\,\)TeV–PeV astrophysical neutrino program, while a new radio array will probe much higher energies. Using Askaryan emission in the ice, IceCube-Gen2 is expected to achieve sensitivity to neutrinos with PeV energies to up to the EeV scale, enabling the first potential detections of cosmogenic and ultra-high-energy neutrinos with an order-of-magnitude improvement over current limits~\cite{IceCube-Gen2:2020qha, IceCube-Gen2:2021rkf}.

\item[$\bullet$] \textbf{TRIDENT:} The TRIDENT project proposes a multi-km$^3$ water-Cherenkov array in the South China Sea, optimized for energies from the sub-TeV scale up to EeV. With an instrumented volume of $\sim 7.5$ km$^3$ and an angular resolution reaching $\mathcal{O}(0.1^\circ)$ at $\sim 100$ TeV, TRIDENT aims to achieve sensitivities that surpass current optical neutrino telescopes for both diffuse fluxes and point-source searches~\cite{TRIDENT:2022hql}.

\item[$\bullet$] \textbf{P-ONE:} The Pacific Ocean Neutrino Experiment (P-ONE) will deploy deep-sea optical modules in the northeast Pacific to target high-energy neutrinos in the $\gtrsim 10$ TeV–multi-PeV range. Its large instrumented volume and long baselines are designed to deliver competitive sensitivities for diffuse fluxes and point-source searches, complementing existing ice and water telescopes~\cite{P-ONE:2020ljt}.

\item[$\bullet$] \textbf{KM3NeT-ARCA:} Located in the Mediterranean Sea, ARCA (Astroparticle Research with Cosmics in the Abyss) is optimized for the TeV–PeV neutrino window, with sensitivity extending to $\sim 30$ PeV. Its design allows sub-degree angular resolution and large effective areas for both track and cascade events, with strong capabilities for diffuse and transient searches~\cite{KM3Net:2016zxf}.

\item[$\bullet$] \textbf{GRAND 200k:} The Giant Radio Array for Neutrino Detection (GRAND) envisions $\sim 200{,}000$ radio antennas deployed worldwide to detect Askaryan and air-shower signals from neutrinos with $E_\nu \gtrsim 10^{17}\,\mathrm{eV}$ up to $10^{20}\,\mathrm{eV}$. GRAND aims for world-leading sensitivity to cosmogenic neutrinos~\cite{GRAND:2018}.

\item[$\bullet$] \textbf{Baikal-GVD}: The Gigaton Volume Detector (GVD), deployed in Lake Baikal, is a modular, $\sim$ 1km$^{3}$-scale Cherenkov neutrino telescope optimized for TeV–PeV astrophysical neutrinos. Its deep-water site provides excellent optical properties (long absorption and scattering lengths), enabling sub-degree angular resolution for track-like events and competitive cascade reconstruction. Baikal-GVD is designed for efficient detection of both diffuse and point-source fluxes. Recent expansions have significantly enhanced its effective area and sensitivity to transient sources, making it a key contributor to the global high-energy neutrino network \cite{Baikal-GVD:2018isr}.

\item[$\bullet$] \textbf{ANITA V:} The proposed fifth flight of the Antarctic Impulsive Transient Antenna (ANITA) will continue the balloon-borne search for impulsive radio signals from neutrino interactions in Antarctic ice. ANITA targets neutrinos in the $\sim 10^{18}$–$10^{21}$ eV range, probing the ultra-high-energy frontier with upgrades in hardware and analysis over earlier flights~\cite{ANITA:2018}.

\item[$\bullet$] \textbf{TAMBO:} The Tau Air-shower Mountain-Based Observatory (TAMBO) is a proposed deep-valley detector in the Andes, designed to detect Earth-skimming tau neutrinos. TAMBO will probe the poorly explored 1–100 PeV energy window, providing high sensitivity in a region of minimal atmospheric background~\cite{TAMBO:2025jio}.

\item[$\bullet$] \textbf{Trinity:} The Trinity concept envisions air-Cherenkov telescopes viewing the horizon to detect upgoing showers from tau neutrinos in the 1 PeV–10 EeV band. This approach bridges the sensitivity gap between IceCube’s optical array and radio-based ultra-high-energy detectors~\cite{Brown:2021lef}.

\item[$\bullet$] \textbf{POEMMA:} The Probe of Extreme Multi-Messenger Astrophysics (POEMMA) consists of two space-based telescopes operating in stereo to detect both fluorescence and Cherenkov signals from ultra-high-energy neutrinos. POEMMA is designed to achieve sensitivity to tau neutrinos above $\sim 2\times10^{16}$ eV, with unique capabilities for transient and target-of-opportunity observations~\cite{Olinto:2021poemma}.

\item[$\bullet$] \textbf{PUEO:} The Payload for Ultrahigh Energy Observations (PUEO) is a next-generation long-duration balloon experiment building on ANITA’s legacy. With phased-array technology, PUEO is projected to achieve world-leading sensitivity above $\sim 1$ EeV, surpassing ANITA’s performance by more than an order of magnitude at $\lesssim 30$ EeV~\cite{PUEO:2021}.

\item[$\bullet$] \textbf{BEACON:} The Beamforming Elevated Array for COsmic Neutrinos (BEACON) is a mountain-top phased radio array concept targeting upgoing tau-induced air showers in the $10^{17}$–$10^{19}$ eV decade. With scalable deployment of hundreds to thousands of antennas, BEACON aims to exceed current EeV neutrino limits and to probe transient sources \cite{Southall:2022yil}.

\item[$\bullet$] \textbf{ARIANNA:} The ARIANNA experiment deploys autonomous radio detectors on the Ross Ice Shelf, sensitive to Askaryan emission from neutrinos with energies $\gtrsim 10^{17}$ eV. Its modular design and advanced triggering strategies, including machine-learning approaches, promise improved diffuse flux sensitivities in the 0.1–10 EeV regime~\cite{ARIANNA:2015}.

\item[$\bullet$] \textbf{Pierre Auger Observatory:} Although primarily designed for ultra-high-energy CRs, Auger also searches for neutrinos through Earth-skimming tau and down-going channels. It currently provides the most stringent limits on diffuse fluxes in the $10^{17}$–$10^{19}$ eV range and continues to perform competitive transient follow-up searches~\cite{Auger:2015}.

\item[$\bullet$] \textbf{HUNT:} The High-energy Underwater Neutrino Telescope (HUNT) is a next-generation deep-sea observatory led by the Chinese Academy of Sciences, designed to detect astrophysical neutrinos above $\sim100$ TeV and identify the sources of high-energy CRs. Planned for deployment in the South China Sea, HUNT aims to instrument a $\sim$30 km$^3$ volume (about thirty times larger than IceCube), using optical modules connected to an undersea network. A prototype array was successfully tested at 1.6 km depth in 2025, marking the first step toward a large volume deep sea telescope in the Northern Hemisphere \cite{Huang:2023HUNT}.

\end{itemize}

\section{Neutrino overdensities}\label{sec:neutrino_overdensity}

Neutrinos must obey Pauli exclusion principle, which limits the maximum neutrino overdensity $\eta$ achievable at different physical scales. The neutrino number density is restricted within the Fermi sphere \cite{Bondarenko:2023ukx}

\begin{equation}\label{eq:pauli}
n_v \leq \frac{V_p}{(2 \pi)^3}, V_p=\frac{4 \pi}{3} p_{v, \max }^3.
\end{equation}

The neutrino energy density can't exceed the critical density of the Universe $\left\langle E_\nu\right\rangle n_\nu \ll \rho_c$, with $\rho_c=3 H_0^2 M_P^2$, which leads to a limit

\begin{equation}
\eta<10^2 \mathrm{eV} /\left\langle E_\nu\right\rangle \lesssim 10^2 \mathrm{eV}/m_{\nu}.
\end{equation}

In the limit of neutrinos being massless, the condition $\rho_{\nu} \leq \rho_{c}$ combined with Eq.~\ref{eq:pauli} leads to the least stringent limit on cosmological scales on the cosmic neutrino background, of

\begin{equation}
 \eta \leq 2 \times 10^4.
\end{equation}
which is the value we quote in the main text. It should be noticed that neutrinos also cluster gravitationally. The clustering induced in galaxies has been discussed in several works \cite{Ringwald:2004np,LoVerde:2013lta,deSalas:2017wtt,Zhang:2017ljh,Mertsch:2019qjv,Zimmer:2023jbb,Holm:2023rml,Worku:2024kwv, Hotinli:2023scz, Zimmer:2024max}, with enhancements at most of a factor of $10$ compared to the cosmological average, for typical halo masses and currently allowed values of neutrino masses.

\end{document}